\documentclass[aps,prb,showpacs,twocolumn]{revtex4}

\usepackage{graphics}
\usepackage{graphicx}
\usepackage{amsmath}
\usepackage{amssymb}

\newcommand{\insertpic}[1]{\scalebox{0.34}{\includegraphics{#1}}}
\newcommand{\insertpictwo}[1]{\scalebox{0.34}{\includegraphics[angle=270]{#1}}}
\newcommand{\insertpictwob}[1]{\scalebox{0.25}{\includegraphics{#1}}}
\newcommand{\bvec}[1]{{\mathbf #1}}

\begin{document}

\title{Sensitivity of the interlayer magnetoresistance of
 layered metals\\
 to intralayer anisotropies}

\author{Malcolm P. Kennett$^{1}$ and Ross H. McKenzie$^{2}$}
\affiliation{$^1$ Physics Department, Simon Fraser University, 8888 University Drive, Burnaby, British Columbia, V5A 1S6, Canada \\
$^2$ Physics Department, University of Queensland, Brisbane 4072, Australia}
\date{\today}

\begin{abstract}
Many of the most interesting 
and technologically important
 electronic materials
 discovered in the past two decades have both
a layered crystal structure and strong interactions
between electrons.
Two fundamental
questions about such layered metals concern
the origin of intralayer anisotropies and
the coherence of interlayer charge transport.
We show that angle dependent magnetoresistance oscillations 
(AMRO) are sensitive to
anisotropies around an intralayer Fermi surface
and can hence be a complementary probe of such anisotropies
to angle-resolved photoemission spectroscopy (ARPES) and
scanning tunneling microscopy (STM). However,
AMRO are not very sensitive to the coherence
 of the interlayer transport
which has implications for recent AMRO experiments on an overdoped cuprate.
\end{abstract}


\maketitle

\section{Introduction}
For elemental metals, such as tin and sodium,
 it is well established that a Fermi 
liquid description is valid.
 Thus the  Bloch wavevector is a good quantum number
for electronic excitations and 
there is a well-defined three-dimensional Fermi
 surface (FS).\cite{Ashcroft} Any variation of
properties over the FS is of secondary interest.
In contrast, many layered metals (e.g., cuprates
and layered manganites)
are distinctly different. Their properties
cannot be described in terms of a Fermi liquid picture.\cite{norman}
Even when one sees signatures of an intralayer
FS, suggesting quantum coherence of excitations
within individual layers, there is controversy
over what length and time scales the electronic
excitations are 
coherent between layers.\cite{millis} Furthermore, in the cuprates 
 properties such as the pseudogap, the quasiparticle spectral weight,
and scattering rate, vary significantly over the intralayer FS.\cite{norman}
 These variations, also seen in ARPES \cite{Damascelli} and STM \cite{shen} 
 may be key to understanding the origin of the superconductivity and the 
unusual properties of the metallic and pseudogap phases.\cite{norman}
Significant anisotropies in quasiparticle weight were also seen recently 
in layered  manganites.\cite{cmrpseudogap} We show here how
the dependence of the interlayer magnetoresistance on the 
the magnetic field orientation is quite sensistive to intralayer anisotropies.

\begin{figure}[htb]
\insertpictwob{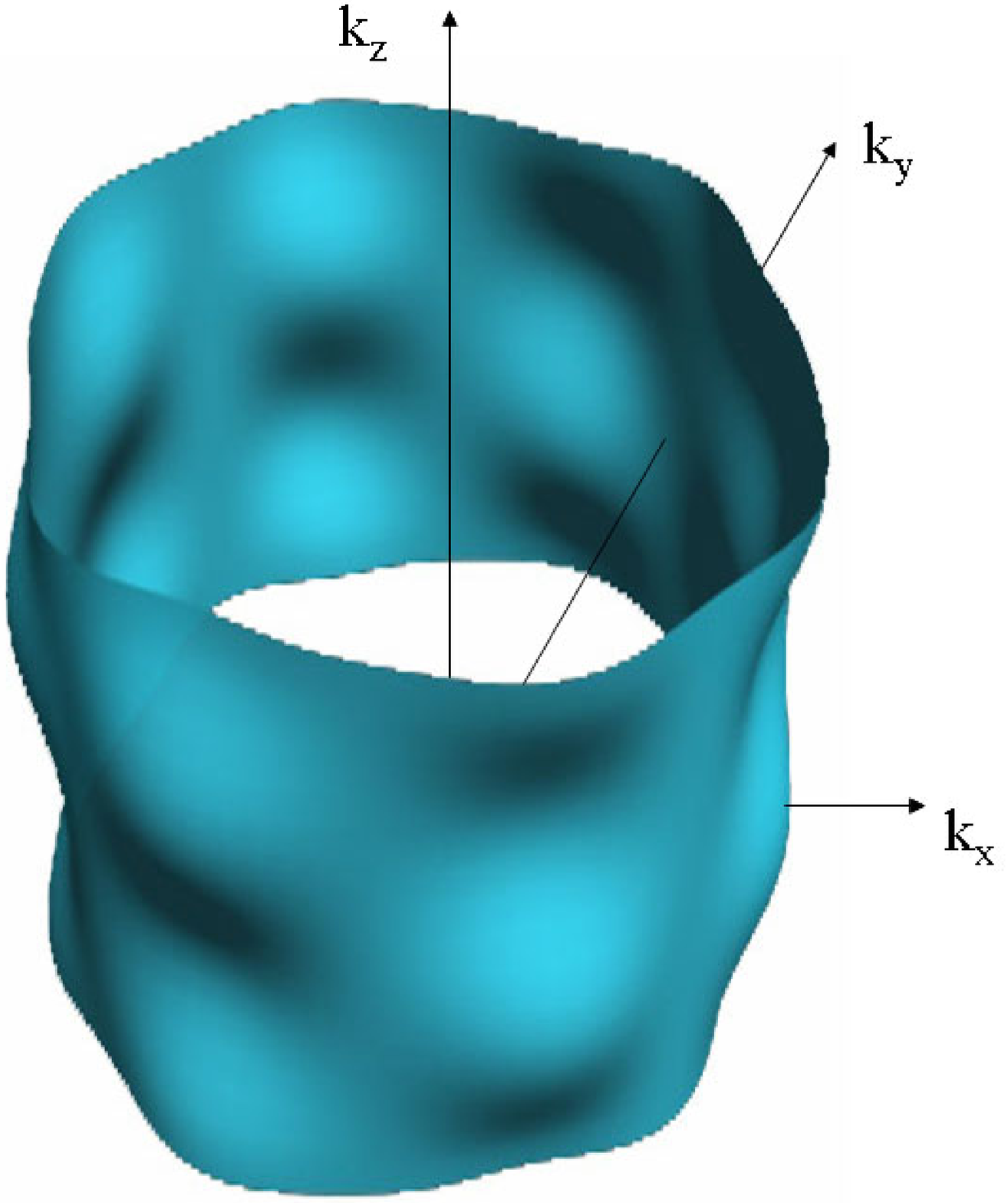}
\insertpictwob{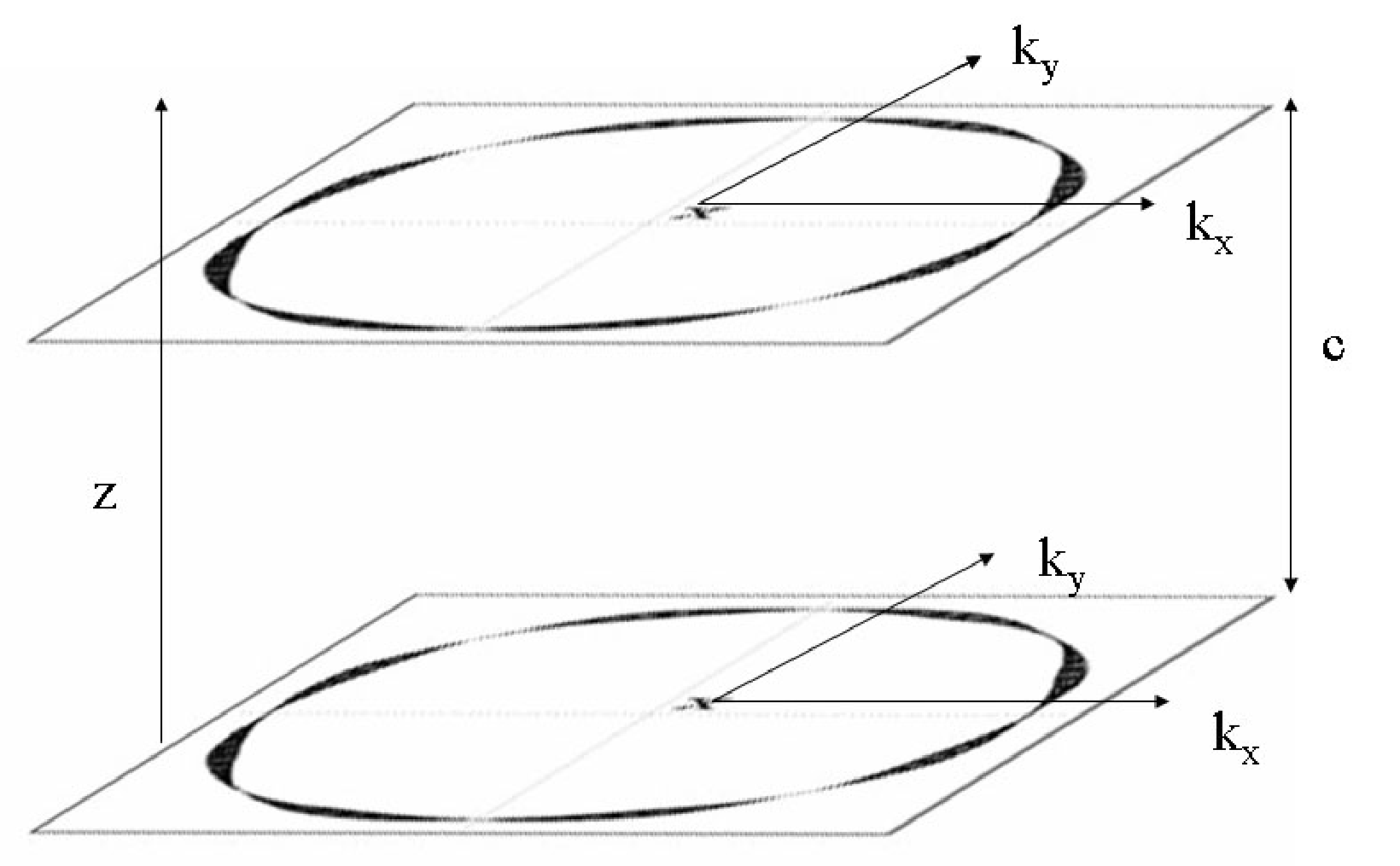}
\caption{Two pictures of interlayer transport.
(i) The Fermi surface (FS) is three dimensional and  warped by
the quantum coherence of electron transport between layers and
the variation  of the interlayer hopping
with the intralayer wavevector. \cite{Nature,Bergemann}
(ii) The FS is only well defined within individual layers.
The interlayer transport is weakly incoherent, i.e. momentum
parallel to the layers is conserved but coherence is only between
neighbouring layers. The thickness of the line is proportional 
to the magnitude of the interlayer hopping. 
} \label{fig:fig1}
\end{figure}

{\it Angle dependent magnetoresistance oscillations (AMRO).}
The dependence of the interlayer magnetoresistance on the
magnetic field direction
has been used to map out a three-dimensional (3d) FS
in a chemically diverse range of layered metals.
including organic charge transfer salts,\cite{kartsovnik-chem,wosnitza}
ruthenates,\cite{Bergemann,Balicas} semiconductor heterostructures,\cite{Osada}
tungsten bronzes,\cite{Beierlein} intercalated graphite,\cite{enomoto}
and an overdoped thallium cuprate.\cite{Nature,Majed}
 However, most of the observed AMRO are also consistent with a 
a two-dimensional FS, i.e., a FS
existing only within the individual layers and weakly incoherent interlayer
transport.\cite{McKMos}
Hence, it is difficult to experimentally distinguish the two
physically distinct pictures summarised in
Fig.~\ref{fig:fig1}.\cite{McKMos}
AMRO are essentially a geometric resonance involving the cyclotron orbits
projected onto the plane parallel to the layers.\cite{CY}  In this sense
they are similar to geometric commensurability oscillations in
magnetoresistance seen for two dimensional electron gases in semiconductor
heterostructures.\cite{Weiss,Beenakker,Kennett}

In this paper we present the following new results for interlayer 
magnetotransport in a layered Fermi liquid metal. 
(i) We give a general expression for the interlayer conductivity
  in a tilted magnetic field including the effects of anisotropy 
around the FS of the Fermi wave vector, Fermi velocity,
scattering rate,  quasiparticle weight, 
  and interlayer hopping integral. 
(ii) We derive (i) for both coherent interlayer transport
  (a three-dimensional FS)
 and for weakly incoherent interlayer transport 
  (where the FS is only well defined within the layers). 
[See Fig.~\ref{fig:fig1}].
 Furthermore, we elucidate the physics behind why
 the same result is obtained 
  for both types of interlayer transport.\cite{foot}
(iii) Our results show that a three-dimensional FS is 
{\it not necessary} to give a quantitative description of AMRO experiments
 on an overdoped cuprate.\cite{Nature}  The data can  be reproduced by 
  the weakly incoherent model. Others have considered the effect on AMRO
 of various specific anisotropies and specific field directions,
\cite{kartsovnik-chem,aniso,Bergemann,Drag}
but not the general case considered here.

The paper is structured as follows.  In Sec.~\ref{sec:anis} we introduce 
the anisotropies in FS properties that we consider and state
our main result, Eq.~(\ref{eq:anisall}) for interlayer conductivity 
in the presence of these anisotropies.  In Sec.~\ref{sec:deriv} we discuss the 
derivation of our main result for the interlayer conductivity for both
coherent and weakly incoherent interlayer transport. 
In Sec.~\ref{sec:exp} we explore the implications of these results 
for AMRO experiments on thallium cuprate.\cite{Nature} Finally, in 
Sec.~\ref{sec:discuss} we conclude and give a discussion of our results.

\section{Anisotropies in the Fermi surface and transport}
\label{sec:anis}
There are a number of FS properties that can be anisotropic in layered
metals.  We discusss intralayer anisotropy in the Fermi surface, 
dispersion and in-plane scattering, and anisotropy in the interlayer 
hopping.  All these anisotropies have been considered in earlier
works, but not simultaneously, which is the generic situation.

We will consider the interlayer conductivity in a tilted magnetic field,
\begin{eqnarray}
\bvec{B} = B
(\sin \theta \cos \varphi, \sin \theta \sin \varphi , \cos\theta),
\label{eq:field}
\end{eqnarray} 
which is at an angle $\theta$ to the $c$-axis and makes an azimuthal angle
$\varphi$ to the $a$-axis in the $ab$ plane
(see inset to Fig.~\ref{fig:fig2}).

\subsection{ Intralayer anisotropies.}

We use the angle $\phi$ to parametrize 
the intralayer FS defined by
\begin{eqnarray}
 \bvec{k}_F(\phi)= k_F(\phi)(\sin \phi, \cos \phi).
\label{eq:fs}
\end{eqnarray} 
For an anisotropic in-plane dispersion, $\epsilon_{2d}(k_x,k_y)$, the
Fermi wavevector $\bvec{k}_F$, which maps out a surface (strictly speaking,
a curve) of constant energy surface, and 
is defined by
$ \epsilon_{2d}(\bvec{k}_F(\phi))= E_F,$
 where $E_F$ is the Fermi energy.  
The Fermi velocity is defined by, 
\begin{eqnarray}
\bvec{v}_F(\varphi) = \frac{1}{\hbar} \left.
\frac{\partial\epsilon}{\partial\bvec{k}}\right|_{\bvec{k}= \bvec{k}_F}
\label{eq:vf}
\end{eqnarray} 
and is always normal to the Fermi surface.
Hence, if there is intralayer anisotropy in the 
dispersion relation there must also be anisotropy
in {\em both}
the Fermi surface and 
Fermi velocity.

{\it Cyclotron frequency.}
The semi-classical equations of motion \cite{Ashcroft}
for an electron moving on the intralayer FS
in a magnetic field with component $B \cos \theta$ perpendicular to
the layers can be be solved give
the variation of the angular speed due to cyclotron
motion 
around the intralayer FS
as
\begin{eqnarray}
\omega_0(\phi) =  eB
 \cos\theta \frac{\bvec{k}_F(\phi)\cdot 
\bvec{v}_F(\phi)}{\hbar k_F(\phi)^2} ,
\label{eq:omega0}
\end{eqnarray}
where $\bvec{v}_F(\phi)$ is the Fermi velocity.
For a circular FS $\omega_0(\phi)$ has a constant value,
$\omega_c \cos\theta$, with $\omega_c=eB/m^*$, the cyclotron frequency.
However, for an anisotropic FS, $\bvec{k}_F$ and $\bvec{v}_F$
 are not parallel and $\omega_0(\phi)$ will vary around the FS.

{\it Scattering rate.}
The variation of the transport lifetime
over the FS is given by $\tau(\phi)$.
The probability of an electron not being scattered in moving 
between two points on the intralayer FS, defined by angles
$\phi_2$ and $\phi_1$, is 
\begin{eqnarray}
G(\phi_2,\phi_1)  = 
\exp \left(-\int^{\phi_2}_{\phi_1} 
\frac{ d\psi}{\omega_0(\psi)\tau(\psi)}\right) .
\label{eq:gphi}
\end{eqnarray} 
In what follows an important quantity is
$P \equiv G(2\pi,0)$,   
the probability that an electron makes a complete orbit
of the intralayer FS without being scattered. 

{\it Interlayer hopping.}
The Hamiltonian for hopping between the layers is:
\begin{eqnarray}
{\mathcal H}_\perp = \sum_{ij} t_\perp(\bvec{r}_i - \bvec{r}_j) 
\left[c^\dagger_i c_j e^{i\Phi_{ij}} + c^\dagger_j c_i e^{-i\Phi_{ij}}\right] ,
\end{eqnarray}
where
$\Phi_{ij} = \left(ec/\hbar\right)
\left(A_z(\bvec{r}_i)- A_z(\bvec{r}_j)\right)$ 
is the Aharonov-Bohm (AB) phase acquired by an electron
 hopping between $\bvec{r}_i$ in one layer  and $\bvec{r}_j$
 in the other layer, \cite{CY} with layer separation $c$ and  
$\bvec{A_\parallel}=A_z(\bvec{r})
 \hat{\bvec{z}}= \frac{1}{2} (\bvec{B_\parallel} \times \bvec{r})$
the vector potential for the magnetic field parallel to the layers,
$\bvec{B_\parallel}$.
Momentum anisotropy in $t_\perp$ is present in the Fourier transform
 of the hopping matrix element
$t_\perp(\bvec{r}_i - \bvec{r}_j)$ via
\begin{eqnarray}
t_\perp(\phi)  & = & \int d^2\bvec{r}
\exp(i\bvec{k}_F(\phi)\cdot\bvec{r}) \, t_\perp(\bvec{r}) .
\label{eq:hopft}
\end{eqnarray}
In momentum space the difference in AB phases acquired in 
hopping between layers 
for positions $\phi_1$ and $\phi_2$ on the FS 
is
\begin{eqnarray}
\Phi(\phi_2,\phi_1) & = & c\tan\theta \left[k_F(\phi_1)
\cos(\phi_1-\varphi) \right. \nonumber \\
& & \left. \hspace*{1.5cm}
 - k_F(\phi_2)\cos(\phi_2 - \varphi)\right] . %\nonumber
\label{eq:phi}
\end{eqnarray} 
We will also see that
for a coherent three-dimensional FS
this quantity can be related to the Bloch wavevector 
perpendicular to the layers (cf. Eq.~(\ref{eq:bloch}) below).

\subsection{The main result}

We now state  a new result, central to this paper.
The interlayer conductivity, $\sigma_c$ in a
tilted magnetic field, [Eq.~(\ref{eq:field})]
at zero temperature is
\begin{eqnarray}
\sigma_{c}(\theta,\varphi) & = & \frac{s_0 eB\cos(\theta)}{(1-P)}
 \int_0^{2\pi} \frac{d\phi_2}{\omega_0(\phi_2)} 
 \int_{\phi_2 - 2\pi}^{\phi_2}
\frac{d{\phi_1}}{\omega_0(\phi_1)} \nonumber \\ & & \times \, t_\perp(\phi_2) t_\perp(\phi_1) 
 \cos\left(\Phi(\phi_1,\phi_2)\right) 
G(\phi_2,\phi_1) , \nonumber \\ & & 
\label{eq:anisall}
\end{eqnarray}
\noindent
where 
$s_0 = (e^2c/\pi^2\hbar^4)$. 
The factor $1/(1-P)$ in the conductivity comes from noting the $2\pi$ 
periodicity of the integrand and reducing the integration region for $\phi_1$.
 The factor
 $e B \cos \theta /\omega_0(\phi_2)$
can also be interpreted as anisotropy in the density of states.\cite{Drag}
Equation~(\ref{eq:anisall}) 
does not explicitly take into account possible anisotropy in the 
quasiparticle weight $Z(\phi)$, due to many-body effects.
  The hopping matrix elements
$t_\perp(\phi)$ should be understood as 
$t_\perp(\phi) = Z(\phi)t_\perp^0(\phi)$, where $t_\perp^0(\phi)$ is
the bare hopping.
It should be stressed that Eq.~(\ref{eq:anisall}) depends {\it only}
on {\it intralayer} FS properties, and
holds irrespective of the
particular form of those anisotropies.

\begin{figure}[htb]
\insertpic{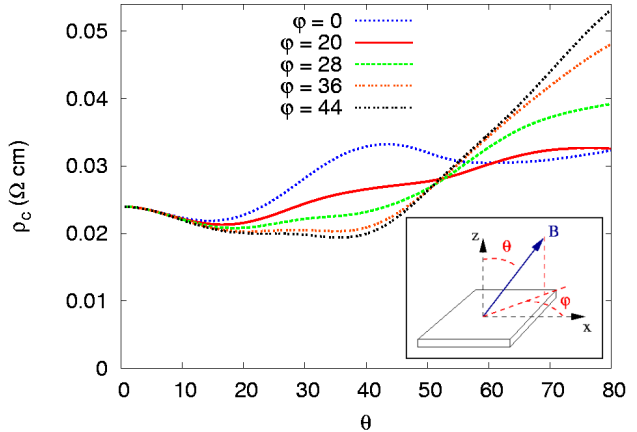}
\caption{Calculated angular dependence of the interlayer resistance for
 for several different azimuthal angles.
The parameters are chosen (see Sec.~\ref{sec:exp} for precise parametrization) 
so that the calculated curves
agree essentially perfectly with the measured curves for
the overdoped cuprate  Tl2201.\cite{Nature}
The inset shows the direction of the magnetic field
relative to the layers of the crystal.}\label{fig:fig2}
\end{figure}

In Sec.~\ref{sec:deriv} we give the derivation of this result for both
 coherent transport perpendicular to the layers and for
weakly incoherent interlayer hopping (see Fig.~\ref{fig:fig1}),
 provided $t_\perp \ll E_F$.
Below, in Fig.~\ref{fig:fig2} we show the calculated
AMRO for parameters that fit the experimental data 
for the overdoped thallium cuprate in
Ref.~\onlinecite{Nature}.
The fitting procedure in Ref.~\onlinecite{Nature}
allowed for anisotropy in the FS, but not
in the Fermi velocity (i.e., in $\omega_0(\phi)$) or
in the intralayer scattering.
Allowing for both of these factors we find a very high quality,
quantitative fit to the
data, which for clarity of presentation is not shown, since it
lies virtually on top of our calculated magnetoresistance.

\subsection{Mapping out Fermi surface anisotropies at high magnetic fields}
We now consider 
AMRO in high magnetic fields ($\omega_{00}\tau_0 \gg 1$),
since such experiments have previously been
used to map out the FS for very
 clean organic materials. \cite{kartsovnik,wosnitza,kartsovnik-chem}
When  $c k_F \tan \theta \gg 1$ and $c{\rm v}_F\tau_0 eB/\hbar^2 \gg 1$ 
the method of steepest descents may be used to
 evaluate the integrals in Eq.~(\ref{eq:anisall}).\cite{peschansky}
The integrals are dominated
by the angles near $\phi_0$ which is where 
the Fermi wavevector, $k_F(\phi)$ has the
maximum projection along the direction of the intralayer magnetic field, 
$\bvec{b}_\parallel\equiv (\cos\varphi,\sin\varphi)$.
$\phi_0$ is a $\varphi$ dependent quantity and is found from the solution
of $$\frac{\partial}{\partial\phi} \left(\bvec{b}_\parallel(\varphi)
\cdot \bvec{k}_F(\phi)\right) = 0.$$
This leads to 
\begin{eqnarray}
\sigma_c(\theta,\varphi) & \simeq & \frac{2\pi s_0}{(1-P)}
\frac{t_\perp(\phi_0)^2}{(\omega_0(\phi_0))^2} 
\frac{eB\cos\theta}{|\mu(\phi_0)|c\tan\theta} 
\label{eq:asympt}
 \\
& & \times
\left[1 + 2P^\frac{1}{2} \sin\left(2c \tan\theta \,  
\bvec{b}_\parallel(\varphi) 
\cdot \bvec{k}_F(\phi_0)\right)\right],    \nonumber 
\end{eqnarray}
where 
$\mu(\phi)= \frac{\partial^2}{\partial\phi^2}[\bvec{b}_\parallel(\varphi)
\cdot \bvec{k}_F(\phi)]$.
This expression shows that 
 the interlayer magnetoresistance
oscillates as a function of the field tilt angle
and
for a fixed value of $\varphi$ the magnetoresistance
will be a maximum when the field is at  angles $\theta_n$
given by\cite{peschansky,kartsovnik,kartsovnik-chem}
\begin{eqnarray}
\label{eq:kartov}
\bvec{b}_\parallel(\varphi) 
\cdot \bvec{k}_F(\phi_0)
c \tan\theta_n = \pi (n - 1/4)
\end{eqnarray}
where $n=1,2,\dots$.
This expression has been used to
map out the intralayer FS for a wide range
of metallic organic charge transfer salts.\cite{kartsovnik,kartsovnik-chem}
We point out that our expression
(\ref{eq:asympt}) will also be sensitive to
angular variations in the interlayer hopping and cyclotron frequency.
Interestingly, when $\omega_{00} \tau_0 \gg 1 $
there is no $\phi_0$ dependence from
the angular variation in the scattering rate since
the only dependence on the scattering rate is through the 
quantity $P $, which involves an average over the FS.
Hence, the
FS shape may be determined  independently of scattering, which 
can then be used determine the anisotropy in scattering at smaller 
fields.\cite{PhysicaB}

\section{Derivations for coherent and incoherent interlayer transport}
\label{sec:deriv}
We now sketch the derivation of Eq.~(\ref{eq:anisall}) for the
cases of coherent and weakly incoherent interlayer transport.

\subsection{Coherent interlayer transport.}
If $t_\perp \gg \hbar/\tau$, 
then there is a well-defined Bloch wavevector perpendicular
to the layers, $k_z$ and the three-dimensional dispersion is 
\begin{equation}
{\mathcal E}_{3d} = {\mathcal E}_{2d}(k_x,k_y) - 2t^0_\perp(k_x,k_y) 
\cos( c k_z) .
\end{equation}
To obtain Eq.~(\ref{eq:anisall}) within a picture of 
coherent interlayer transport one starts from the dispersion, which 
can be used to determine the Fermi velocity, 
and semi-classical equations of motion for motion on the FS.
Solution of the appropriate
Boltzmann equation, to leading order in $(t_\perp/E_F)^2$,
leads to Eq.~(\ref{eq:anisall}) as a generalization of the
Shockley-Chamber's (SC) formula. \cite{Ashcroft,footnote2}

A useful relation between the projection of the motion in real space
onto the plane of the layers and $c$-axis momentum
 (where $\bvec{R}_\parallel$ is the position in the plane) 
comes from considering the equation of motion 
 (we neglect higher order terms in $t_\perp/E_F$).
Integrating to get  in-plane and $c$-axis momentum gives
\begin{eqnarray}
\bvec{k}_\parallel(t) & = &
 \bvec{k}_\parallel(0) + \frac{eB_\perp}{\hbar}
\left[\bvec{R}_\parallel(t) - \bvec{R}_\parallel(0)\right]
\times \hat{\bvec{z}}   , \label{eq:kparallel} \\
k_z(t) & = &  k_z(0) + \frac{\bvec{B}_\parallel \cdot (\bvec{k}_\parallel(t) 
- \bvec{k}_\parallel(0))}{B_\perp}  . \label{eq:kperp}
\end{eqnarray}
Hence, the interlayer velocity is 
\begin{eqnarray}
v_z(\phi) & = &
\frac{2ct^0_\perp(\phi)}{\hbar}
 \sin \left( ck_z(0) + ck_F(\phi)\tan\theta\cos(\phi-\varphi) \right) .
\nonumber \\ &&
\label{eq:bloch}
\end{eqnarray}
The SC formula \cite{Ashcroft} involves correlations
in this velocity at different times (equivalently different $\phi$)

\begin{eqnarray}
\sigma & = & \frac{e^2}{4\pi^3} \int d^3\bvec{k}_1 v_z(\bvec{k}_1) 
\left(-\frac{\partial f_T}{\partial \epsilon}\right)
\nonumber 
\\ & & \times \int^{\phi_1}_{-\infty}
d\phi_2 \frac{v_z(\phi_2)}{\omega_2(\phi_2)} G(\phi_2,\phi_1) ,
\label{eq:scformula}
\end{eqnarray}
where $G(\phi_2,\phi_1)$ was defined in Eq.~(\ref{eq:gphi}), 
$f_T$ is the Fermi-Dirac distribution, and the limits of integration
for $k_z$ are between $\pm \frac{\pi}{c}$.
The expressions in Eqs.~(\ref{eq:kparallel})-(\ref{eq:bloch})
(when the integral over $k_z$ is performed in Eq.~(\ref{eq:scformula}))
allow us to see how the term associated with an  Aharonov-Bohm phase for weakly 
incoherent transport (c.f., Eq.~(\ref{eq:phi})) arises for coherent 
interlayer transport.\cite{CY}

\subsection{Weakly incoherent interlayer transport.}
 The interlayer current
at $\bvec{r}_1$ from layer 2 is \cite{Mahan}
\begin{eqnarray}
j_\perp(\bvec{r}_1) &=& ie\int d^2\bvec{r}_2 \, t_\perp(\bvec{r}_1-\bvec{r}_2)
\left[c^\dagger_2(\bvec{r}_2)c_1(\bvec{r}_1) e^{i\Phi_{12}} - h.c.
\right] , \nonumber 
\end{eqnarray}
where $\Phi_{12}$ is the gauge phase for hopping between the layers.
Note that $t_\perp(\bvec{r}_1,\bvec{r}_2)$ contains 
the variation of the interlayer hopping over the FS
[Eq.~(\ref{eq:hopft})].

  In linear response the Kubo formula 
for the interlayer conductivity at zero temperature is \cite{Mahan,McKMos}
\begin{eqnarray}
\sigma_c    & = & \frac{2ce^2}{\hbar L_x L_y} {\rm Re}\left[
 \int d^2\bvec{r}_1 
\int d^2\bvec{r}_2
\int d^2\bvec{r}_3 \int d^2\bvec{r}_4 \right. 
\label{eq:inc-cond}
 \\ & & \left.
 \times \,t_\perp(\bvec{r}_1,\bvec{r}_4)t_\perp(\bvec{r}_3,\bvec{r}_2)
e^{i(\Phi_{12} - \Phi_{34})}
K_{E_F}(\bvec{r}_1,\bvec{r}_2,\bvec{r}_3,\bvec{r}_4) \right] 
, \nonumber 
\end{eqnarray}
where \cite{Rudin}
$K_\epsilon(\bvec{r}_1,\bvec{r}_2,\bvec{r}_3,\bvec{r}_4)  =  
G^R_{2,\epsilon}(\bvec{r}_1,\bvec{r}_2)
G^A_{1,\epsilon}(\bvec{r}_3,\bvec{r}_4)$
and $G^R_{1,\epsilon}$ is the retarded 
one-electron Green's function
within layer 1.
Now, $K_\epsilon$ oscillates rapidly as a function of the spatial co-ordinates, with a period comparable to the Fermi wavelength. 
We can separate the non-oscillatory part by introducing co-ordinates
$\bvec{R}_1 = \frac{1}{2}(\bvec{r}_1 + \bvec{r}_4)$, and $\bvec{R}_2 =
 \frac{1}{2}(\bvec{r}_2 + \bvec{r}_3)$ and write

\begin{eqnarray}
K_\epsilon(\bvec{r}_1,\bvec{r}_2,\bvec{r}_3,\bvec{r}_4) & = & \int 
\frac{d^2\bvec{k}_1}{(2\pi)^2}\int \frac{d^2\bvec{k}_2}{(2\pi)^2}
K_\epsilon(\bvec{k}_1,\bvec{R}_1;
\bvec{k}_2,\bvec{R}_2) \nonumber \\ & & \times
e^{i\bvec{k}_1\cdot(\bvec{r}_1 - \bvec{r}_4) 
+ i\bvec{k}_2\cdot(\bvec{r}_3 - \bvec{r}_2)} .
\end{eqnarray}
We use a semi-classical approximation since  $K_\epsilon$ will be
sharply peaked at energies around  the Fermi energy, which
enforces  the condition that the magnitude of
the momentum in the Fourier transform is $\phi$-dependent and
 lies on the FS and then
$K_\epsilon(\bvec{k}_F(\phi_1),\bvec{R}_1;\bvec{k}_F(\phi_2),\bvec{R}_2)$
 satisfies the Boltzmann-type equation for a diffuson:\cite{Rudin}

\begin{eqnarray}
\left[\bvec{v}_F(\phi_2)\cdot\frac{\partial}{\partial \bvec{R}_2} +
\omega_0(\phi_2)\frac{\partial}{\partial\phi_2} 
- \frac{1}{\tau(\phi_2)}\right]
 K_{E_F}(\phi_1,\bvec{R}_1;\phi_2,\bvec{R}_2) 
 \nonumber \\
 =   
 2\pi \frac{\omega_0(\phi_1)}{eB\cos\theta} 
\delta(\phi_1-\phi_2)\delta(\bvec{R}_1 - \bvec{R}_2) , \nonumber 
\end{eqnarray}
where the factor of $\omega_0$ on the right hand side of the equation comes
from noting that the Boltzmann equation is initially stated with 
$\delta^2(\bvec{k}_1 - \bvec{k}_2)$ on the right hand side.
We can solve this equation for $K$ by
Fourier transforming to get a solution in
terms of 

\begin{eqnarray}
K(\phi_1, \bvec{R}_1; \phi_2, \bvec{q}) & = & \int \frac{d^2\bvec{q}}{(2\pi)^2}
e^{i\bvec{q}\cdot\bvec{R}_2} K(\phi_1, \bvec{R}_1; \phi_2, 
\bvec{R}_2), \nonumber \\ & &
\end{eqnarray}
and then

\begin{eqnarray}
K(\phi_1, \bvec{R}_1; \phi_2, \bvec{q}) & = & 2\pi e^{-i\bvec{q}\cdot
\bvec{R}_1}\int_{-\infty}^{\phi_2}d\tilde{\phi} \, \frac{\delta(\phi_1 - 
\tilde{\phi})}{\omega_0(\tilde{\phi})} 
\frac{\omega_0(\phi_1)}{eB\cos\theta} \nonumber \\
& & \times e^{-\int_{\phi_2}^{\tilde{\phi}} 
\frac{d\psi}{\omega_0(\psi)\tau(\psi)}} e^{i\bvec{q}\cdot
\int_{\phi_2}^{\tilde{\phi}} \frac{\bvec{v}_F(\psi)}{\omega_0(\psi)}d\psi} , 
\nonumber \\ & &
\end{eqnarray}
where we can note that
$$\int_{\phi_2}^{\tilde{\phi}} \frac{\bvec{v}_F(\psi)}{\omega_0(\psi)}d\psi
= \bvec{R}_\parallel(\phi_2) - \bvec{R}_\parallel(\tilde{\phi}).$$
Inserting this solution in Eq.~(\ref{eq:inc-cond}), 
using Eq.~(\ref{eq:kparallel}) and integrating over
 $\bvec{R}_1$ and $\bvec{R}_2$ forces
\begin{eqnarray}
\Phi & = & \Phi_{12} - \Phi_{34} =  c\tan\theta (k_F(\phi_2)
\cos(\phi_2-\varphi) \nonumber \\  & & \hspace*{4cm}
 - k_F(\phi_1)\cos(\phi_1 - \varphi)). \nonumber 
\end{eqnarray}
so we get as our final result
\begin{eqnarray}
\sigma_{c} & = & \frac{2ce^2}{\hbar L_x L_y} {\rm Re}
\int_0^{2\pi} d\phi_2 \int_{-\infty}^{\phi_2} d\phi_1 \int d\bvec{R}_1 \int
d\bvec{R}_2  \frac{eB\cos\theta}{\omega_0(\phi_1)}
\nonumber \\ & & \times \frac{eB\cos\theta}{\omega_0(\phi_2)}
t_\perp(\phi_1) \, t_\perp(\phi_2) \cos[\Phi] \,
K_\epsilon(\phi_1,\bvec{R}_1;\phi_2,\bvec{R}_2), \nonumber \\ & &
\end{eqnarray}
which is identical to the equation found with the Boltzmann equation 
for coherent transport in Eq.~(\ref{eq:anisall}) after using periodicity of
the integrand to reduce the interval of integration over $\phi_1$ to have
length $2\pi$.

\section{AMRO in thallium cuprates}
\label{sec:exp}
In Secs.~\ref{sec:anis} and \ref{sec:deriv} we established Eq.~(\ref{eq:anisall})
for general forms of anistropy and interlayer transport mechanisms.  
To illustrate the use of Eq.~(\ref{eq:anisall}) and its significance we fit
recent AMRO measurements of a thallium cuprate.\cite{Nature,Majed}
An important point to note about these experiments is that the data is very
high quality and it is thus possible to fit AMRO to very high precision, 
which leads to extraction of FS parameters to high accuracy.

\begin{figure*}[htb]
\insertpictwo{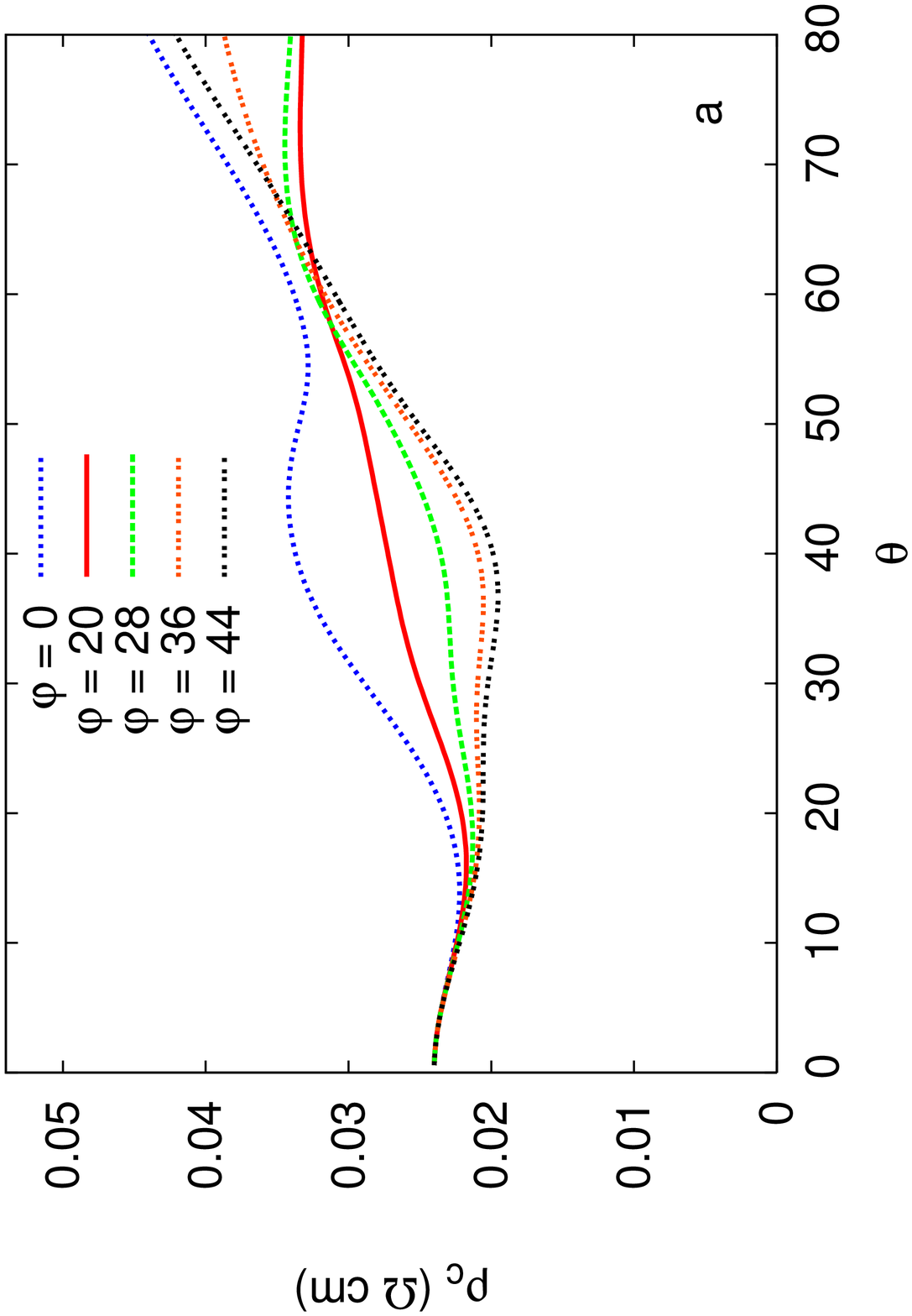}
\insertpictwo{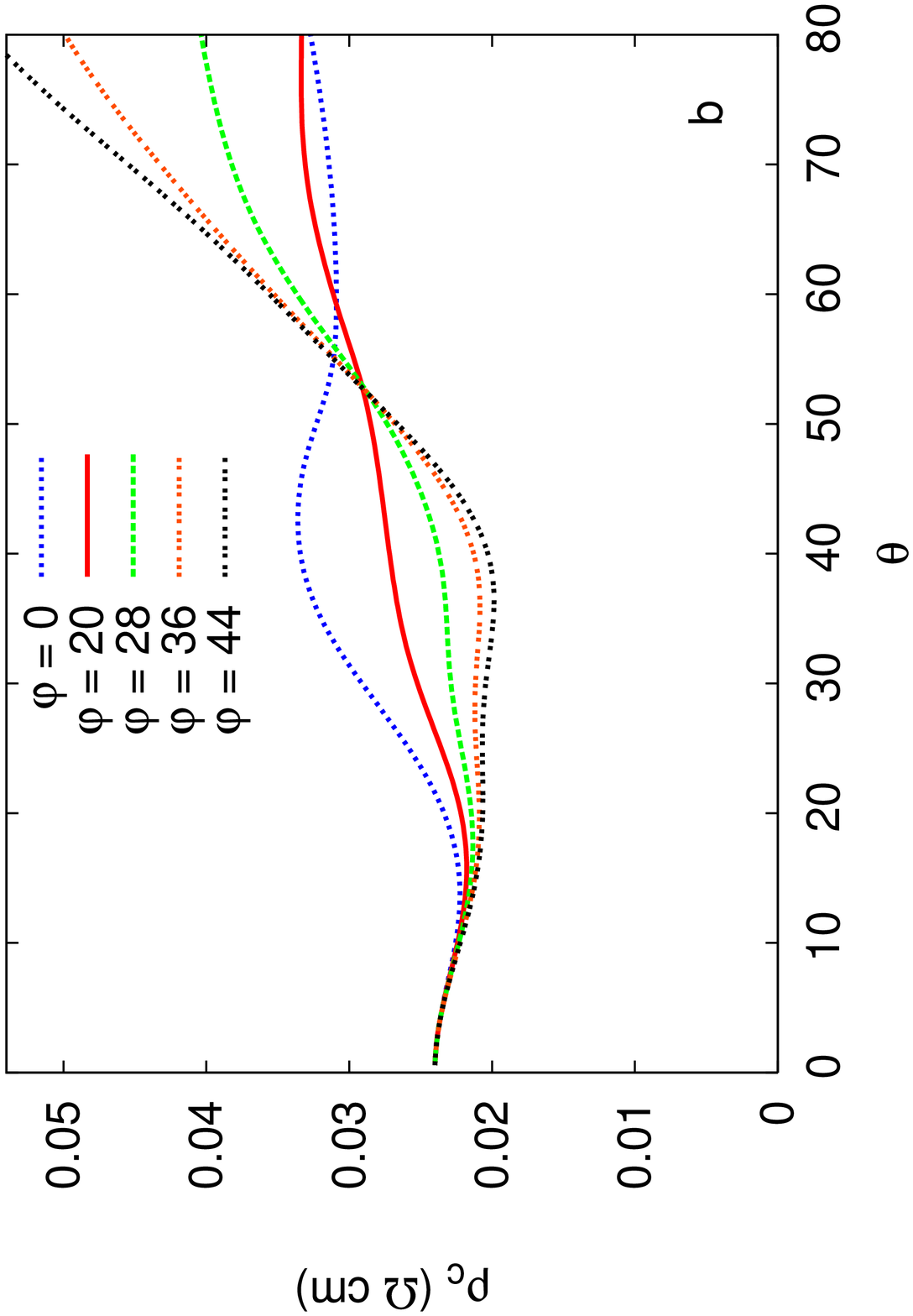}
\insertpictwo{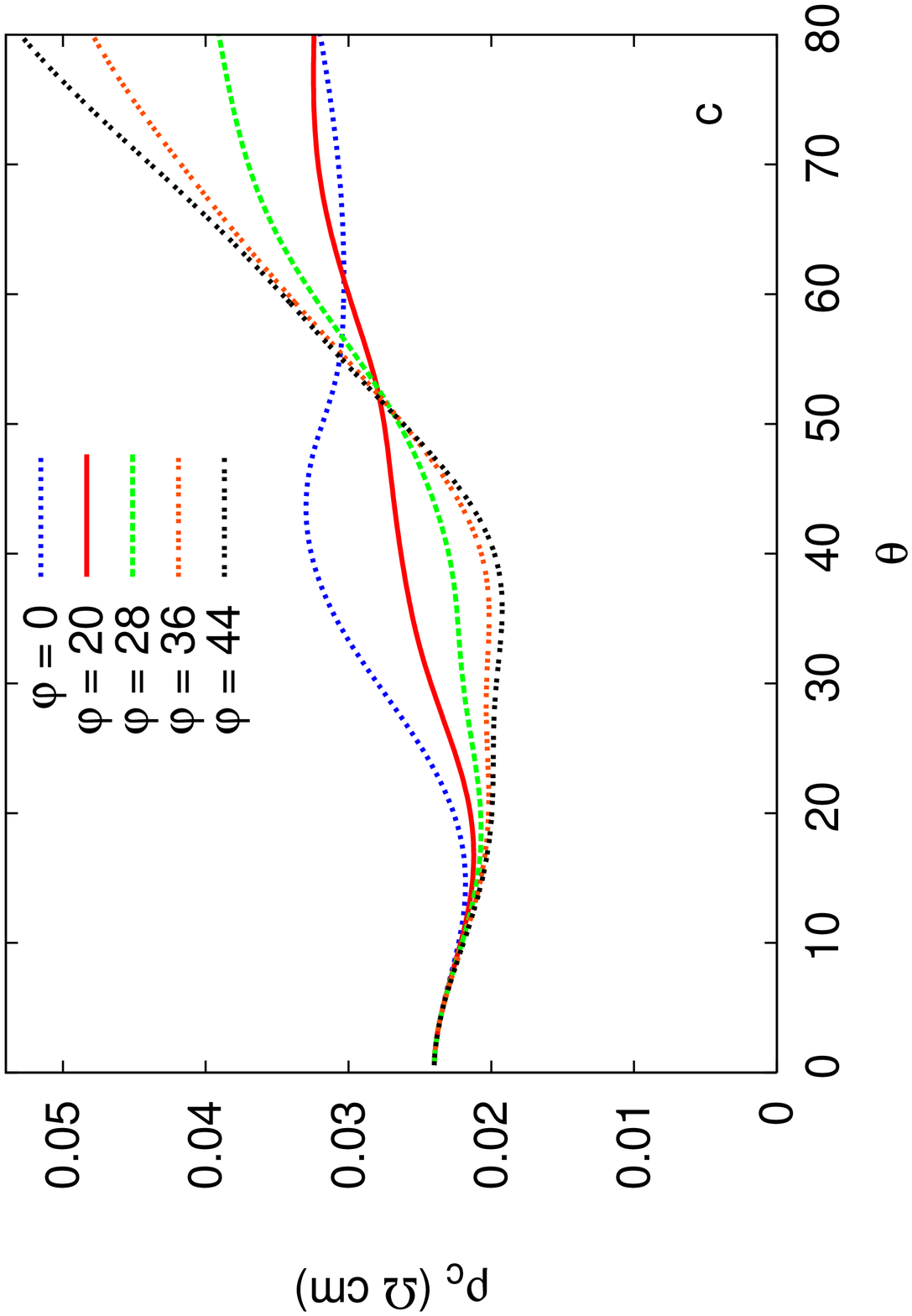}
\caption{Calculated angular dependence of the interlayer magnetoresistance
for several different azimuthal angles,
based on Eq.~(\ref{eq:anisall}).
For the parameter values used,
$\eta_1 = 0.675$, $t_\perp^0 \simeq 13 \ {\rm meV}$,
 $c k_F^0 = 8.64$, $\omega_{00} \tau_0 =0.45$.
In a) $\alpha =  0.0$, $\kappa = 0.0$, and
$u = 0.0$, as defined in Eqs.~(\ref{eq:tperp2})
to  (\ref{eq:aniskF}).  In b) we allow anisotropy in $k_F$, with
$\kappa = -0.033$.  In c) we allow anisotropy in $k_F$ and $\omega_0$
with $\kappa = -0.033$, and $u = - 0.08$.}
\label{fig:addpara}
\end{figure*}

Numerical evaluation of Eq.~(\ref{eq:anisall}) shows that
changing the functional forms of the
anisotropy in interlayer hopping and intralayer
scattering leads to quantitative variations in the AMRO.
This can be seen in Figs.~\ref{fig:addpara} a-c
as we add in anisotropies to our fitting of AMRO.
To extract deviations of the FS from circularity
in a self-consistent manner from AMRO experiments
 one must also take into account the related anisotropy
in the Fermi velocity and the resulting anisotropy in the
angular speed (c.f., Eq.~(\ref{eq:omega0}).
 In previous work, \cite{Bergemann,Nature}
such effects were not included in the expressions used
for fitting experimental data.  This omission quantitatively
changes the magnitude of anisotropies in $k_F$ extracted
from the fits. Similarly, a self-consistent determination of anisotropy
in $\tau(\phi)$ also requires including anisotropy in $\omega_0(\phi)$.

\subsubsection*{Model forms for anisotropy in thallium cuprates}
We use a form of $t_\perp(\phi)$
which is consistent with the body-centred-tetragonal
crystal structure of thallium cuprate
and band structure calculations:\cite{vdMarel,Nature}
\begin{equation}
t_\perp(\phi) = t_\perp\left(\sin(2\phi) + \eta_1 \sin(6\phi) + \eta_2
\sin(10\phi)\right),
\label{eq:tperp2}
\end{equation}
parametrized by $\eta_1$ and $\eta_2$.
The crystal symmetry requires that \cite{Bergemann}
$ t_\perp(\phi) = -t_\perp(\phi + \pi/2) $,
implying $ \eta_1 = 1 + \eta_2$, and that
$ t_\perp(\phi)$ has eight-fold symmetry and vanishes
at $\phi = \frac{n\pi}{4}$, $n=0,1,..,7$.

A number of experiments on the cuprates suggest that
the scattering rate has a four-fold variation around the Fermi surface.
Such variations are also seen in calculations for a doped Hubbard
model, using Cellular Dynamical Mean-Field theory.
\cite{Civelli} 
A simple hot or cold spots model of scattering \cite{Ioffe,Sandeman} 
yields
\begin{eqnarray}
\frac{1}{\tau(\phi)}
 & = & 
\frac{1}{\tau_0}
\left(1 + \alpha\cos(4\phi)\right)  ,
\label{eq:anistau}
\end{eqnarray}
where $\alpha > 0$ for cold spots located
 at $\phi = \pm \frac{\pi}{2}, \pm \frac{3\pi}{2}$.
We parametrize the anisotropy in the FS and the cyclotron
frequency as:
\begin{eqnarray}
k_F(\phi) &  = & k_F^0\left(1 + \kappa\cos(4\phi)\right) ,  \\
\frac{1}{ \omega_0(\phi)}      
 & = & 
\frac{1}{\omega_{00}}
\left(1 + u \cos(4\phi)\right).
\label{eq:aniskF}
\end{eqnarray}

In Fig.~\ref{fig:addpara} we show the effects of adding various
anisotropies to the calculated AMRO.  In Fig.~\ref{fig:addpara}a we
show the calculated AMRO only allowing for anisotropic interlayer hopping.
This demonstrates that much of the overall form of the magnetoresistance curves
 is determined by the presence of of  the eight nodes in the interlayer 
hopping [Eq.~(\ref{eq:tperp2})].  Due to this symmetry we only
display AMRO for $\varphi \in [0,\frac{\pi}{4}]$.
In Fig.~\ref{fig:addpara}b we add the effects of an anisotropic $k_F$ which can
be seen to depress $\rho_c(\theta)$ for small $\varphi$ at large $\theta$ 
and enhance $\rho_c(\theta)$ for larger $\varphi$ (up to $\frac{\pi}{4}$)
at large $\theta$.  In Fig.~\ref{fig:addpara}c we make our fit self-consistent 
by also allowing for anisotropy in $\omega_0$, which tends to reduce 
$\rho_c(\theta)$ at larger $\theta$ for larger $\varphi$.
When we also include scattering anisotropy, we recover the fit that we
displayed in Fig.~\ref{fig:fig2} for the following set of parameters:
$\eta_1 = 0.675$, $t_\perp^0 \simeq 13 \ {\rm meV}$,
 $c k_F^0 = 8.64$, $\omega_{00} \tau_0 =0.45, \kappa = -0.033, u = - 0.08$,
and $\alpha = 0.01$.

The effect of anisotropic
scattering on the shape of the
curves is most pronounced at large values of $\theta$, and
the height of the peak in $\rho_c(\theta)$ at  
$\theta \sim 40^\circ$, for $\varphi \sim 0^\circ$.  We use a small value
of $\alpha$ here to fit the data, but larger values tend to be required
at higher temperatures.\cite{Majed}
Increasing $\alpha$ also increases the value of $\rho(\theta = 0) = \rho_0$.

For the overdoped cuprate we consider,
a tight-binding model for the intralayer bands
has been fit to ARPES data for a sample of similar
chemical composition.\cite{Plate}
We use this model as an independent check of the reliability
of the values we obtain for $k_F^0$, $\omega_0$, $u$, and $\kappa$.
The tight-binding fit gives comparable values
which increases confidence that the extraction
of anisotropies from AMRO is robust.

Equation (\ref{eq:anisall}) is valid for
both coherent and weakly incoherent transport between layers, except
in a small region near $\theta = 90^\circ$.
 For coherent transport, but not for incoherent transport,
 there is a peak in the resistivity at $\theta = 90^\circ$ due to  orbits
on the $3d$ FS that do not exist if one can only
define a $2d$ FS within the layers.\cite{kartsovnik-chem,McKMos}
A second definitive signature
of  coherent interlayer transport is beats in quantum magnetic oscillations
\cite{McKMos} but these have not been seen in thallium cuprate.
For the samples in Ref.~\onlinecite{Nature},  $t_\perp$ and
$\hbar/\tau_0$, are of the same order,
which would place them at the boundary
between coherent and incoherent interlayer transport. However,
$\omega_{00}\tau_0$ is insufficiently large to
be able to see either of
these definitive signatures of coherence.  This implies that
further evidence is required to validate the claim of a 3$d$ FS
in Ref.~\onlinecite{Nature}, and so a 3$d$ FS is {\it not} necessary to
explain the AMRO data in Ref.~\onlinecite{Nature}.

\section{Discussion}
\label{sec:discuss}
In conclusion, we have given a general formula for AMRO in layered
metals which have anisotropy in interlayer tunnelling, in intralayer
scattering, and an anisotropic FS.  We have performed
explicit calculations for parameters relevant to experiments on
thallium cuprate \cite{Nature} and shown that these do not on their own 
imply the coherence of interlayer transport. 
These fits are very sensitive to the anisotropies of intralayer properties.

Whilst we have focused our attention on the example of AMRO in thallium
cuprate, our results have a much wider applicability -- in fact to any
layered metal with an anisotropic Fermi surface and anisotropic scattering.
Our results have been stated for quasi-two-dimensional systems, but it should be
fairly straightforward to generalize 
our results to quasi-one-dimensional systems
such as Bechgaard salts,
 for which there is a substantial amount of AMRO
 data.\cite{wosnitza,kartsovnik-chem}
  We also discuss how AMRO can be used in conjunction with other 
techniques such as ARPES to place strong constraints on FS properties.  
Combined AMRO and ARPES studies of layered metals could give a strong
consistency check for the results of both techniques and this suggests 
many future opportunities ahead for AMRO as a powerful probe of 
anisotropies in layered metals.
In particular, for superconducting organic 
charge transfer salts 
a characterization of intralayer anisotropies,
could reveal the presence of a pseudogap with d-wave symmetry,
as predicted by a resonating
 valence bond theory of these materials.\cite{powell}

\acknowledgements

We thank N. E. Hussey and M. Abdel-Jawad for stimulating discussions and
for showing us unpublished experimental results.
We thank B. J. Powell for a critical reading of the manuscript.
This work was supported by the ARC (R.H.M.), and NSERC (M.P.K.).
We thank Wolfson College, Oxford for hospitality.

%%%%%%%%%%%%%%%%%%%%%%%%%%%%%%%%%%%%%%%%%%%%%%%%%%%%%%%%%%%%
%
%   References.
%
%%%%%%%%%%%%%%%%%%%%%%%%%%%%%%%%%%%%%%%%%%%%%%%%%%%%%%%%%%%%


\begin{thebibliography}{99}
\bibitem{foot}
It is also worth noting that
 the derivation given here for weakly incoherent transport is 
actually simpler than
that given previously for the more restrictive
case of isotropic intralayer Fermi surface properties.  \cite{McKMos}
\bibitem{footnote2} Note that $t_\perp/E_F \sim 0.03$ for the sample in
Ref.~\onlinecite{Nature}.
\end{thebibliography}
\end{document}